\documentclass[preprint,12pt]{elsarticle}

\usepackage{amsmath}
\usepackage{amssymb}
\usepackage[most]{tcolorbox}
\usepackage{ulem}
\newcommand\hl{\bgroup\markoverwith{\textcolor{yellow}{\rule[-.5ex]{.1pt}{2.5ex}}}\ULon}
\usepackage{calc}

\begin{document}

\begin{frontmatter}

\title{Symmetric Nash equilibrium of political polarization in a two-party system}

\author{Jonghoon Kim\fnref{fn1}}
\address{Department of Physics, Pukyong National University, Busan 48513, Korea}
\fntext[fn1]{Current address: Department of Energy Engineering, Korea Institute of Energy Technology, Naju 58330, Korea}
\author{Hyeong-Chai Jeong}
\address{Department of Physics and Astronomy, Sejong University, Seoul 05006,
Korea}
\author{Seung Ki Baek\corref{cor2}}
\address{Department of Scientific Computing, Pukyong National University, Busan 48513, Korea}
\ead{seungki@pknu.ac.kr}

\begin{abstract}
The median-voter hypothesis (MVH) predicts convergence of two party platforms
across a one-dimensional political spectrum during majoritarian elections. From
the viewpoint of the MVH, an explanation of polarization is that each election
has a different median voter so that a party cannot please all the median voters
at the same time. We consider two parties competing to win voters along a
one-dimensional spectrum and assume that each party nominates one candidate out of two
in the primary election, for which the electorates represent only one
side of the whole population. We argue that all the four candidates will come to
the same distance from the median of the total population through best-response
dynamics.
\end{abstract}


\end{frontmatter}

\section{Introduction}

Statistical physics deals with macroscopic patterns emerging from
microscopic interactions. Collective decision-making is an
example of such emergent phenomena arising from individual-level
interactions~\cite{arganda2012common,bhattacharya2014collective,hartnett2016heterogeneous,de2017criticality}.
Broadly speaking, the group of activities associated with collective
decision-making in
society can be referred to as politics, and it is why statistical physicists
have viewed the interplay between individual choices and political changes
within the framework of complex
systems~\cite{galam2021will,baek2020co,bottcher2020competing,cardoso2020gender}.
One of widely accepted political values is
democracy, according to which the people have the right to rule.
A democratic
government thus has an excellent incentive to meet the people's demands, but the
price is political instability: the governmental policies may suddenly change
if the power is transferred to the opposing party by election,
and the change can be especially drastic when political opinions are
polarized to a great degree.

Whether political polarization is an inevitable part of democracy is not clear.
As will be explained below,
in a two-party system, the median-voter hypothesis (MVH) predicts that the two
parties along a one-dimensional political spectrum will converge to the median
voter's position through majoritarian elections~\cite{downs1957economic}.
The assumptions of the MVH might look too restrictive, but it has been reported
in congressional voting that very often the issue reduces to a
one-dimensional matter~\cite{mccarty2019polarization}.
In such a one-dimensional political landscape, a two-party
system is robust, as assumed by the MVH,
because a centrist party cannot easily find a political
``niche'' between two existing parties~\cite{macarthur1967limiting}. Thus,
the question still remains: Why are they so polarized?

It may be attributed to social homophily or echo
chambers~\cite{mcpherson2001birds,madsen2018large}, and temporary polarization
may be consistent with Bayesian updating~\cite{dixit2007political}.
Many of the existing approaches assume that people update
their political positions through mechanistic interaction with neighbors such as
homophily, assimilation, and
differentiation~\cite{levin2021dynamics,axelrod2021preventing,leonard2021nonlinear,macy2021polarization}.
A recent study answers this question by pointing out that voters are not as
rational as assumed in the MVH~\cite{yang2020us}: According to this idea,
voters will rather {\it satisfice} than maximize their utility functions. By
adapting to such voting behavior with a simple gradient ascent method, the
parties can develop symmetric polarization, maintaining an equal distance from
the median voter.

In this work, we wish to examine an alternative
explanation~\cite{campbell2018polarized}, which argues that the median voter's
position experienced by a party may be different from election to election:
For example, a party has to nominate a candidate before a presidential election.
To become the nominee, candidates in the primary should take
opinions from their own supporters seriously, even if the supporters' overall
position differs from the general public opinion, because their
votes are needed to win the primary.
By considering the nomination process as a part of election, this approach
explains permanent polarization without resorting to cognitive biases and
bounded rationality.
In addition, our work provides a testable prediction that
polarization will only increase further if the payoff for
the loser of the final election also becomes valuable enough,
which has not been proposed by other models.

In the next section, we will formulate our model as a game of
two-round competition among four players, whose strategies are their political
positions. The Nash equilibria will be identified.
We then propose how to update the players' positions to reach one of the Nash
equilibria. Such dynamic consideration is important in several aspects: First,
the existence of plausible dynamics
ensures the feasibility of the discovered equilibrium. Second, the time scale
to reach the equilibrium can be estimated: If it diverges, deviation from the
equilibrium can prove more pervasive than predicted by a static
analysis~\cite{kim2019sex,pangallo2019best}. Third, when multiple equilibria
exist, the distribution of convergence points will generally require dynamical
consideration.
As a model of dynamics, one could consider an evolutionary process which works
in a large population of simple-minded agents,
but we believe that a variety of strategic moves in the
course of election can be better described by best-response dynamics among the
candidates, according to which each candidate deliberately seeks a position to
be the final winner given that the other competitors' positions are held fixed.
The whole dynamics is thus projected onto four candidates, and the
electorate react to the candidates instantaneously with fixed political
positions, which is an assumption that we have borrowed from the MVH.
The key point is that everyone in our model society seeks
the best response to the given configuration, and this is one of common
approaches to model human behavior in evolutionary game
theory~\cite{fudenberg1998theory}. In other words,
each player tries to maximize his or her own objective function, as is
not uncommon in physics, but an important difference of game theory from
physics is that such individual optimization may drive the total system away
from an optimal point.

\section{Model}

\subsection{Median-voter hypothesis}
\label{sec:mvh}

The MVH is based on two main assumptions~\cite{mueller2003public}:
First, political positions
are defined along a one-dimensional spectrum. Second, each voter has single-peaked
preferences along the axis. To define single-peakedness, let $U_i(x)$ be
voter $i$'s utility function defined over $x$.
The ideal point for $i$, denoted as $x_i^\ast$, is such that $U(x) <
U(x_i^\ast)$ for any $x \neq x_i^\ast$. Choose two points $y$ and $z$ along the
$x$ axis so that either $y,z > x_i^\ast$ or $y,z < x_i^\ast$.
Voter $i$'s preferences are single-peaked if and only if $U_i(y) >
U_i(z)$ is equivalent to $|y-x_i^\ast| < |z-x_i^\ast|$. That is, when $y$ and
$z$ are on the same side of $x_i^\ast$, voter $i$ prefers $y$ to $z$ if and only
if $y$ is closer to $x_i^\ast$ than $z$ is.

Imagine a set of ideal points of $N$ voters, $\{x_1^\ast, x_2^\ast,
\ldots, x_N^\ast\}$. Let $N_L(x)$ be the number of $x_i^\ast \le x$ for a
certain position $x$, and let $N_R(x)$ be the number of $x_i^\ast \ge x$.
Then, a median position is any number $x_m$ such that $N_L(x_m)
\ge N/2$ and $N_R(x_m) \ge N/2$.
When the above two main assumptions hold true, the MVH states that
this median position does not lose under majority rule.
The reason is the following: For any
$z<x_m$, voters on the right-hand side of $x_m$ prefer $x_m$ to $z$ because of
their single-peaked preferences, and the number of such voters is greater than
or equal to $N/2$ by definition. A similar argument shows that $x_m$
does not lose to any $z>x_m$.

The above analysis on the median voter has provided the basis for
Hotelling's law, which says that two party platforms
will eventually converge to the median voter's position.
This is better described as a working hypothesis
rather than a mathematical statement
because the assumptions behind the MVH are only approximate in reality.
Some researchers have reported that local governments with two
parties tend to adopt more moderate policies than those with single parties, in
support of the MVH~\cite{kasper1971political}, but whether the relationship
between the median voter and policy variables is statistically significant
remains in question~\cite{boyne1987median}.

\subsection{Semifinalists' dilemma}
\label{sec:semi}

\begin{figure}
{\centering
\includegraphics[width=0.4\textwidth]{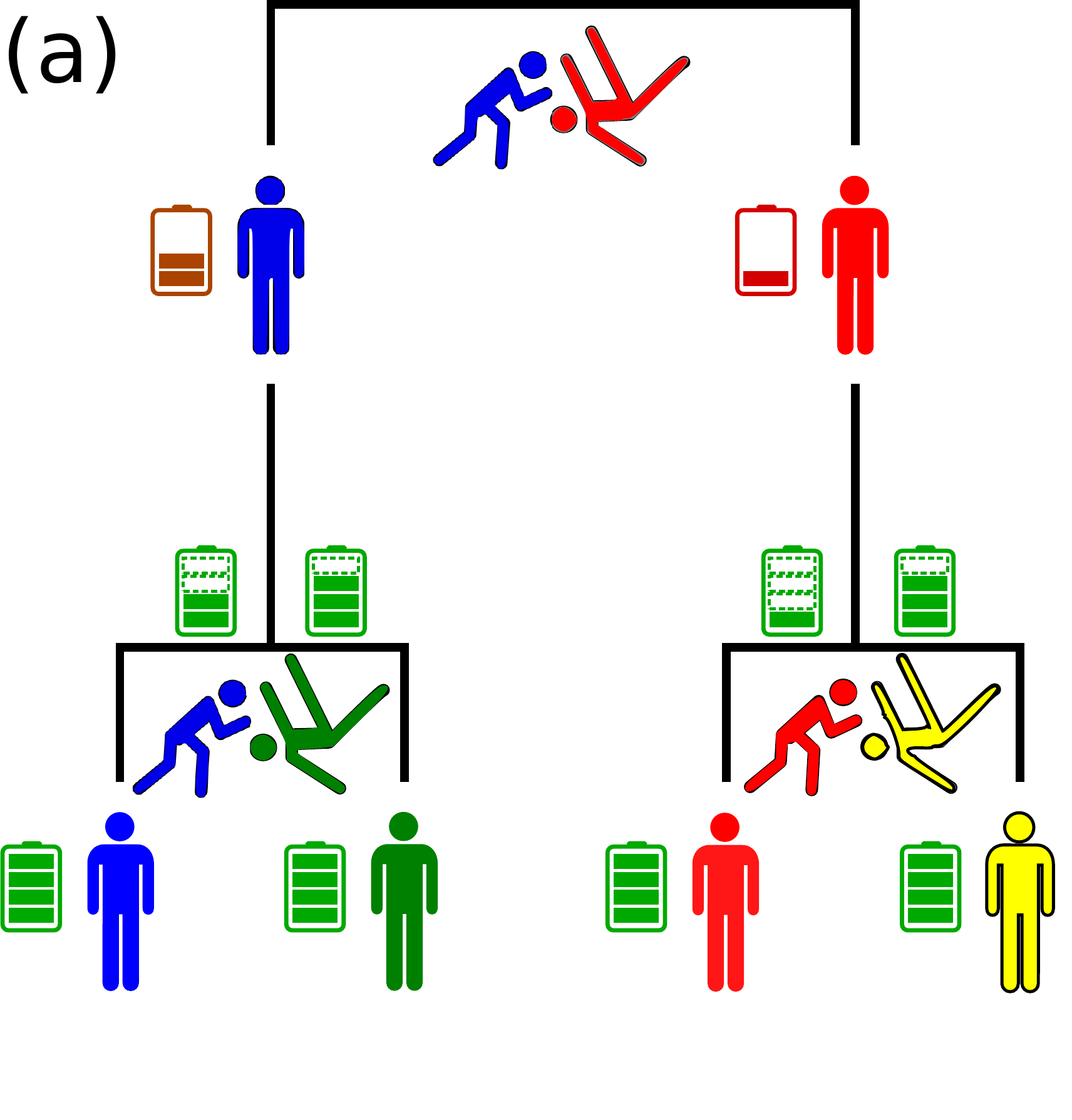}
\includegraphics[width=0.59\textwidth]{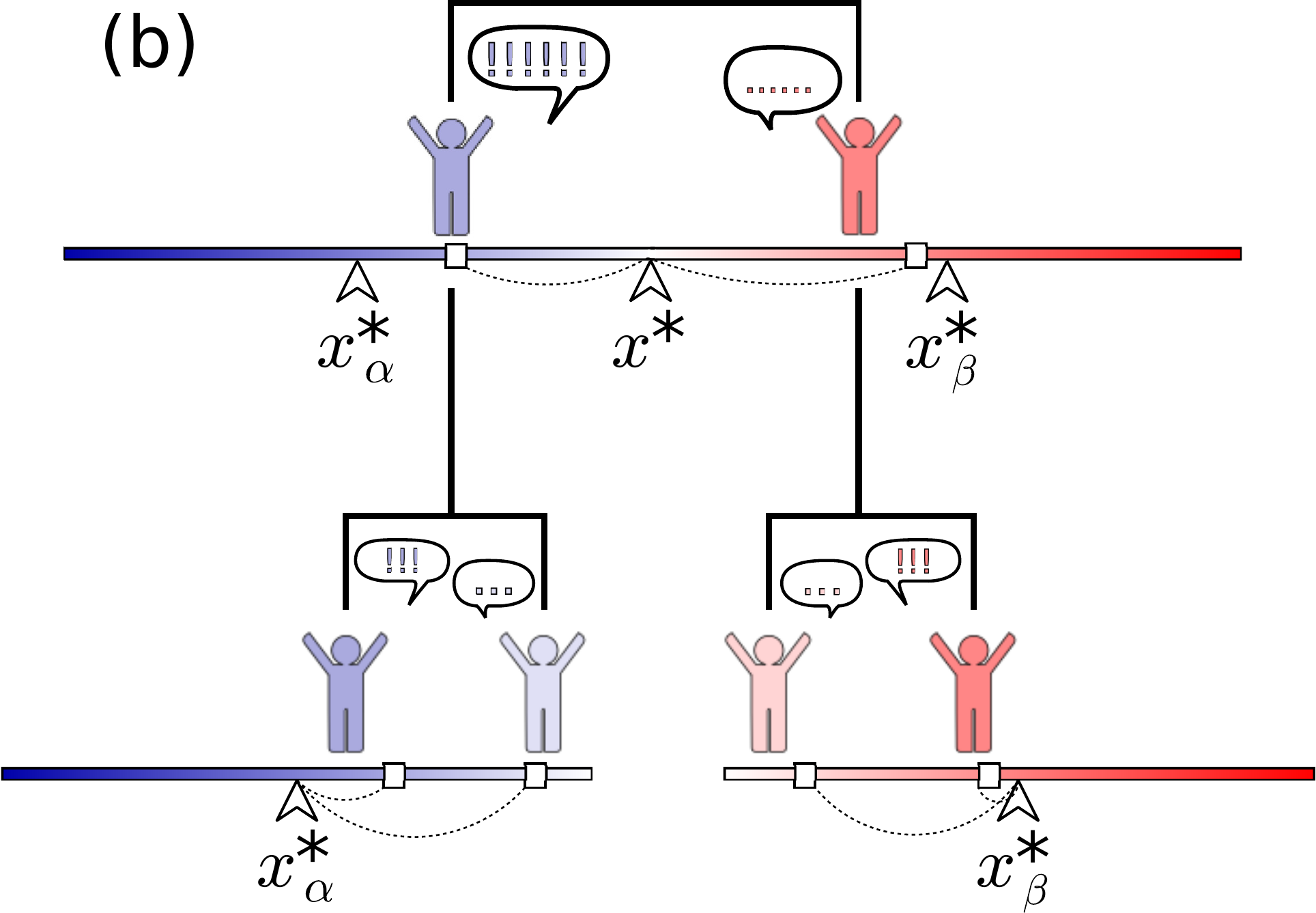}
\par}
\caption{
Schematic diagrams to show
the semifinalists' dilemma and its equivalence to the
two-round election.
(a) In the semifinalists' dilemma, four
players begin with equal stamina (bottom). The one who invests more stamina than
the opponent will proceed to the final, but one cannot win the
final if he or she spent too much stamina in the
semifinal.
(b) In the two-round election, we have two parties,
$\alpha$ (blue) and $\beta$ (red), and each party
has two candidates in the
primary (bottom).
The bluish color box represents the political spectrum among
$\alpha$-supporters, $I_\alpha = [-2,0]$, and the reddish one represents its
$\beta$-counterpart, $I_\beta = [0,2]$.
The candidates' positions are marked in the color boxes by
sliders.
The arrows below the color boxes show the party supporters' median positions,
$x_\alpha^\ast = -1$ and $x_\beta^\ast = 1$, respectively.
Each party organizes a primary
election, in which the
party supporters nominate
one of the candidates for the final by measuring the distances (dotted lines).
The final winner is the one closer to the
position $x^\ast$ of the median voter among the whole electorate.
}
\label{fig:semi}
\end{figure}

Before explaining our model of election,
let us first review a strongly related model called semifinalists'
dilemma~\cite{baek2015nash}:
Imagine a single-elimination tournament among four players, in which only the
final winner earns a unit payoff.
Four equally strong semifinalists have to decide how much stamina to use in the
semifinals, given that only the rest is available in the final
[Fig.~\ref{fig:semi}(a)].
The strategy of a player, say, $i$, is thus the amount of stamina to invest in
the semifinal, which may be regarded as a continuous variable $x_i$
in the unit interval $[0,1]$.
The simplest rule for deciding the winner of a match between two competitors
is that it is the one who invests more stamina in the match.
Therefore, player $i$ must spend enough stamina $x_i$ to pass the semifinal,
and also keep $1-x_i$ large enough to win the final.
The question is how much is the optimal level of $x_i$.

It turns out that this game has infinitely many Nash equilibria. That is,
as long as everyone chooses the same strategy $x$, no one has reason to
deviate from it, regardless of the specific value of $x$: Assume that
player $i$ finds everyone else using a certain common strategy $x$. If $x_i
< x$, player $i$ will lose the semifinal. If $x_i > x$, player $i$
will lose the final. Only by choosing $x_i = x$, player $i$ can ensure
victory with probability $1/4$, which is a natural consequence of symmetry
among the four players.

We may also think of variants of this game, e.g., by assigning a positive
payoff $u<1$ to the runner-up as well. This makes the final less important
because winning the semifinals already guarantees the second place.
If player $i$ spends all the stamina in the semifinal, the expected gain is
$u$, whereas the player would earn $(1+u)/4$ on average by conforming to the
common strategy $x$. The former expected payoff exceeds the latter when
$u>1/3$. Therefore, if the second-place prize is more valuable than
one third of the grand prize, the final will lose attraction, and
players will devote all their efforts to the semifinals.

\subsection{Two-round election}

Let us consider a process of election between two parties, $\alpha$ and $\beta$,
as depicted in Fig.~\ref{fig:semi}(b).
The election consists of two rounds: In the primaries, each party starts with
two candidates and nominates one of them so that the nominees compete in the
second round. By assumption, infinitely many voters are uniformly distributed
inside a one-dimensional interval $I \equiv [-2,2]$. Half of them in
$I_\alpha \equiv [-2,0]$ support $\alpha$, whereas the other half in $I_\beta
\equiv [0,2]$
support $\beta$.
Each of the $\alpha$-candidates, denoted by $A$ and $a$,
chooses a position inside $I_\alpha$, and the same is true for the other party
$\beta$ with candidates $B$ and $b$. The four candidates' positions are their
strategies and will be denoted by $x_A$, $x_a$, $x_B$, and $x_b$, respectively.
As assumed in the MVH, every voter votes for a candidate with the closest
position. The winner of this game is the candidate who wins both the rounds by
getting a larger share of voters than the competitor in each round.

In the primaries, the median voter for $\alpha$ is located at position
$x_\alpha^\ast \equiv -1$, whereas the median voter for $\beta$ is at
$x_\beta^\ast \equiv +1$. In the second round, the median
voter is found at $x^\ast = 0$. Therefore, for any $\alpha$-candidate,
a position outside $\tilde{I}_\alpha \equiv [-1, 0]$ is unreasonable,
and both $x_A$ and $x_a$ must thus belong to $\tilde{I}_\alpha$. By the same
token, both $x_B$ and $x_b$ must belong to
$\tilde{I}_\beta \equiv [0, 1]$.
The connection to the semifinalists' dilemma is clear: Just as a player's
stamina is divided between the semifinal and the final, a candidate's
political position divides an interval of unit length into two pieces.
One of them shows the distance from the median opinion of
party supporters, and the other shows the distance from the origin.
If the former distance is short, he or she has a high chance to win the primary,
but that can be disadvantageous in the final because the median voter at the
origin will not choose such a one-sided position.

It is again
straightforward to show that any symmetric configuration such that $-x_A =
-x_a = x_B = x_b \in [0,1]$ constitutes a Nash equilibrium: Suppose that $A$
tries moving closer to the origin when the four players have reached the
equilibrium.
This move will only make the candidate lose the primary because his or her competitor $a$'s
position is closer to the median voter's position $x_\alpha^\ast = -1$.
On the other hand, if $A$ moves away from the
origin, $A$ will be defeated by the opponent from $\beta$ in the second round
because $\left\| x_A \right\| > \left\| x_B \right\| = \left\| x_b \right\|$.

\section{Results}

\subsection{Numerical simulation of best-response dynamics}

\begin{figure}
\includegraphics[width=0.49\textwidth]{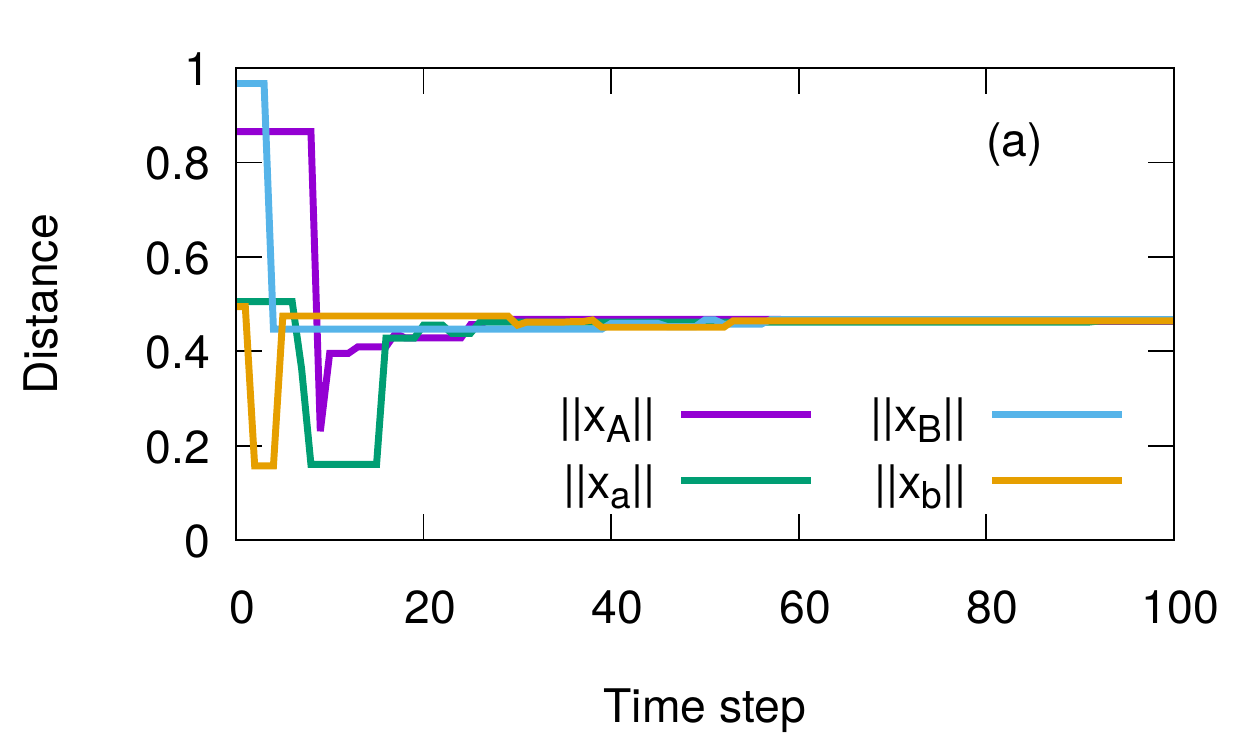}
\includegraphics[width=0.49\textwidth]{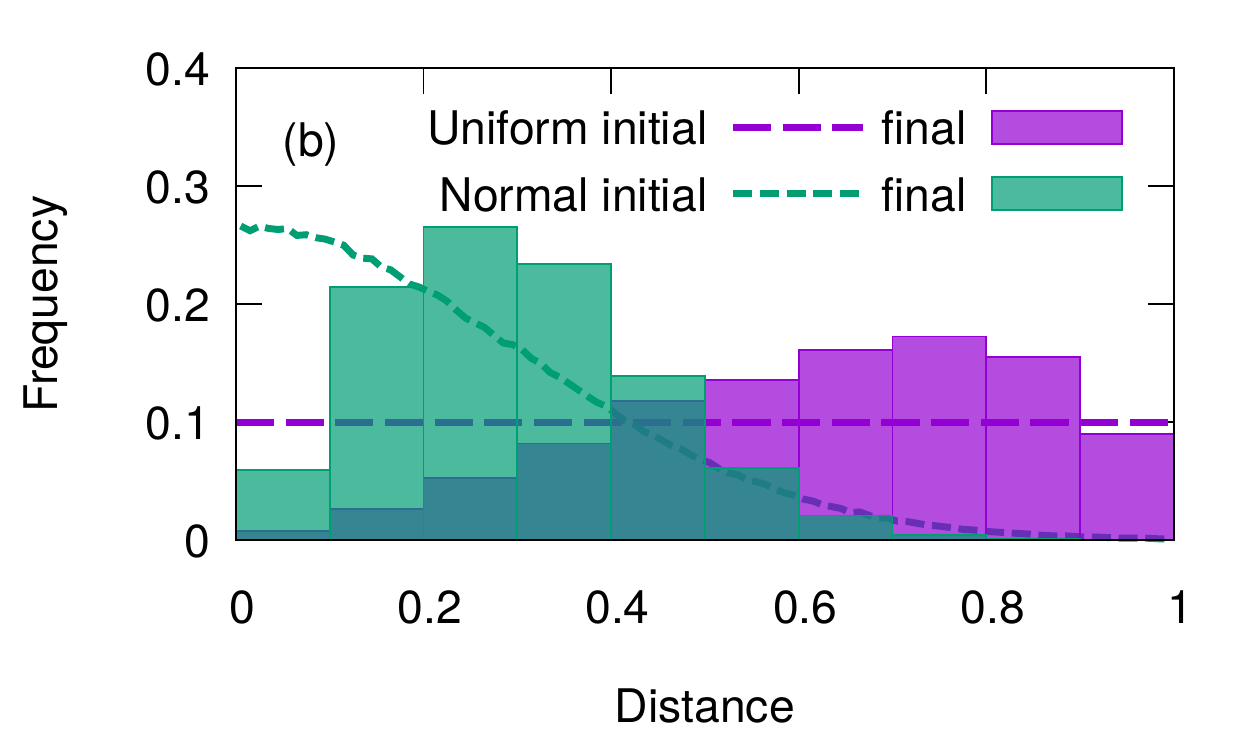}
\caption{(a) A simulation of best-response dynamics, in which the candidates' positions
converge to a Nash equilibrium, which has $\left\| x \right\| \approx 0.46$ in this particular sample. The convergence point depends on the initial positions, the order of updates, and the $M$ trials for updating each candidate's position.
(b)
Histograms of the equilibrium distances from the origin over random initial
conditions and random sequential updates.
The initial positions are drawn either
uniformly from the respective intervals for the parties (purple), or according to the
truncated normal distribution in $[-1,1]$ with a mean at $x=0$ and a scale
parameter $\sigma = 0.3$ (green).
The dark green rectangles represent the overlap between the two histograms.
}
\label{fig:simul}
\end{figure}

To simulate best-response dynamics, we consider the following stochastic dynamics: We have four candidates denoted as $A$, $a$, $B$, and $b$, respectively. The former two, $A$ and $a$, choose their initial positions randomly from $\tilde{I}_\alpha = [-1,0]$, whereas the latter two, $B$ and $b$, do the same from $\tilde{I}_\beta = [0,1]$.
At each time step, let us randomly pick up a candidate $i \in
\{A, a, B, b\}$. Then, if this candidate $i$'s belongs to
party $\alpha$, his or her new position is sampled $M$ times within
$\tilde{I}_\alpha$. Or, if $i$ belongs to $\beta$, the new position is
sampled $M$ times within $\tilde{I}_\beta$. If the new position leads to
$i$'s victory under the assumption that the other three positions are fixed, the
focal candidate $i$ moves to the new position.
If more than one out of the $M$ sampled positions are found to bring victory, the last one will be chosen as $i$'s position at the next time step.
If none of the $M$ sampled positions produce victory, he or she maintains
the {\it status quo}. After trying $M$ new positions for $i$ in this way, the next time step
starts by picking up another candidate from $\{A, a, B, b\}$.
This simulation will reproduce best-response dynamics in the limit of $M \to \infty$,
and we set $M$ to be sufficiently large number of $O(10^2)$ in practice.
This algorithm is statistically equivalent to choosing the first winning position as long as the payoff is binary, i.e., either victory or defeat. We can also speed up the simulation by noting that one cannot gain victory if his or her competitor in the same party has a larger distance from the origin than the candidates of the other party.

Our simulation shows that this dynamics converges to a symmetric Nash
equilibrium within a finite time scale [Fig.~\ref{fig:simul}(a)]. In fact, the
convergence slows down as the distances among the candidates approach $1/M$, but
we may neglect this later phase of convergence in the limit of $M \to \infty$.
Because of the stochasticity of the dynamics, the convergence point depends on the initial positions, the order of updates, and the $M$ trials for updating each candidate's position.
If we take independent samples by generating different random number sequences
uniformly from $[0,1]$, the probability density of $\left\| x \right\|$ after convergence has
a maximum around $0.7$ [Fig.~\ref{fig:simul}(b)]. It is intuitively plausible
that the candidates will not easily converge to $\left\| x \right\| = 0$ or $1$,
but they tend to exhibit a higher degree of polarization than $1/2$,
which is their average distance from the origin at the beginning. The increase of
polarization is a general tendency: We have also tried a unimodal distribution
centered at $x=0$, and the average distance from the origin after reaching a Nash equilibrium is again
greater than at the beginning.

\subsection{Convergence to the Nash equilibrium}

\begin{figure}
{\centering
\includegraphics[width=0.6\textwidth]{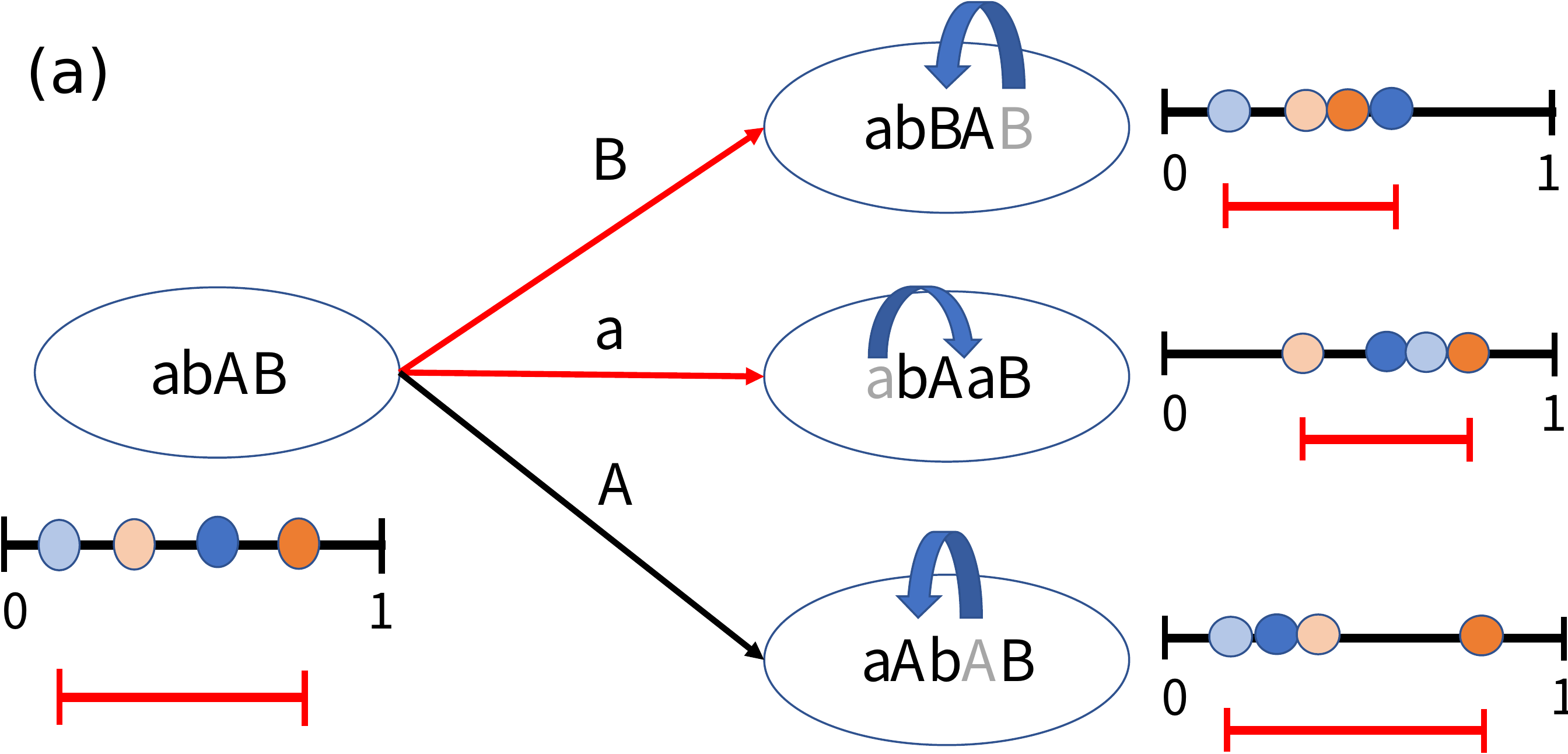}
\includegraphics[width=0.8\textwidth]{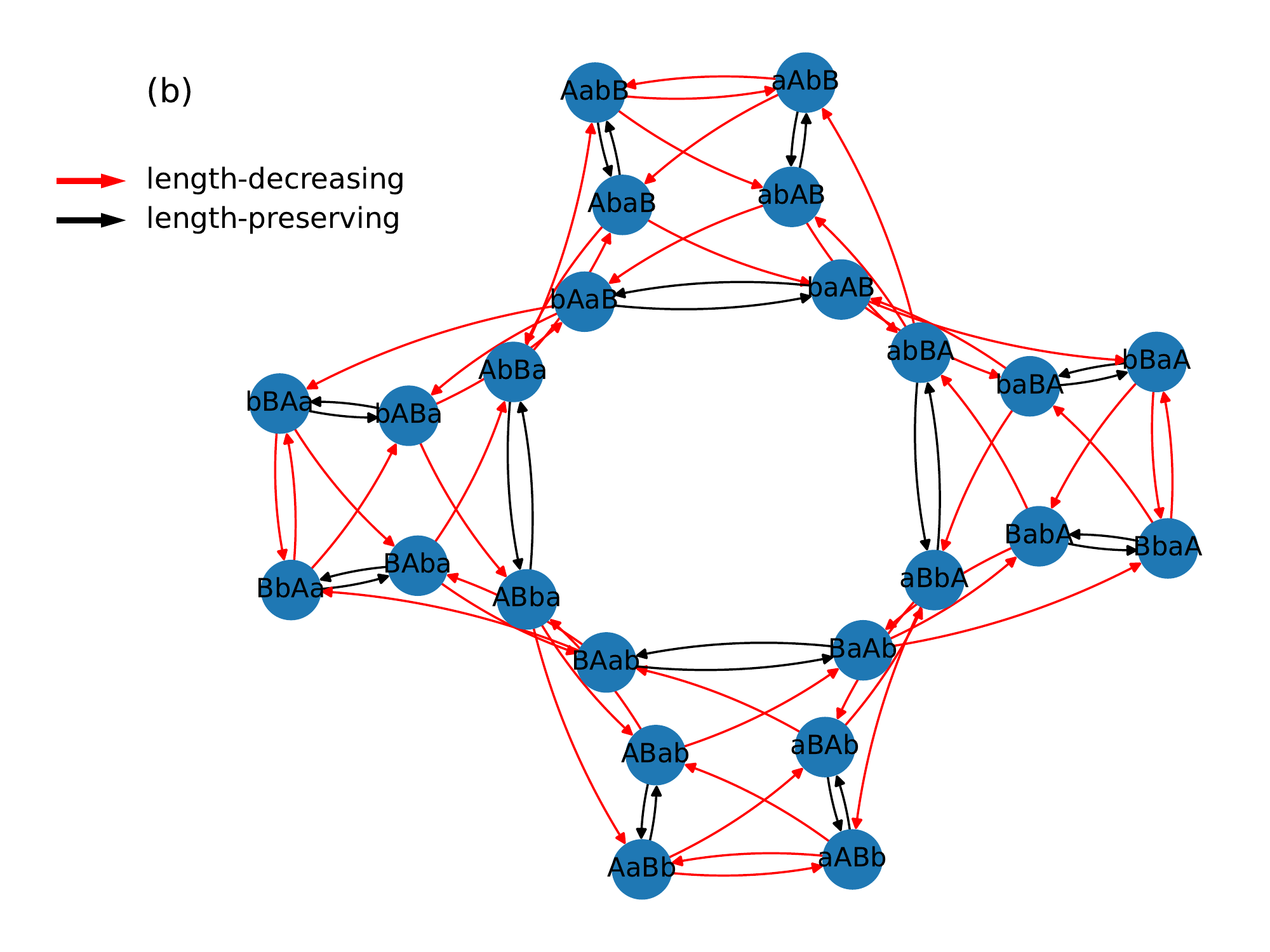}
\par}
\caption{
(a) Elementary transition among states, where a state is represented
by a list of candidates in ascending order of their distances from the origin.
The first transition shows a case that
$B$ is chosen for update, and his or her new position must be between
$b$ and $A$ to defeat both of them.
Or, if it is $a$'s turn as represented by the second transition,
his or her new position must be between $A$ and $B$.
These two transitions, marked as the red arrows, make their positions more
symmetric with respect to the origin than before by shrinking the effective
length $L$ defined in Eq.~\eqref{eq:L}. By contrast, if $A$ is chosen
to move, the transition (black
arrow) preserves $L$.
Note that $b$ has no reason to move because it is
impossible to win this game no matter where he or she goes, and we have omitted
the resulting self-loop.
(b) Combination of elementary transitions
among $24$ possible states. The red and black arrows represent elementary
transitions decreasing and preserving $L$, respectively.
In principle, each node has four incoming links and four outgoing ones, but we have simplified the diagram by omitting a self-loop at each node.
}
\label{fig:transition}
\end{figure}

We can prove the convergence property as follows: Let us write the candidates in
ascending order of their distances from the true median voter at the origin, regardless of their signs. In
Fig.~\ref{fig:transition}(a), we show an example of $abAB$ on the left, which means
that $a$ is the closest candidate to the origin whereas $B$ is the farthest.
From each candidate's point of view, the winning condition is to have a larger
distance than the intra-party competitor to win the first round, but closer than
at least one competitor from the other party to win the second round. If $B$ is
chosen to move in this example, therefore, $B$ has to take a position between
$b$ and $A$. Or, if it was $a$'s turn, $a$ would move somewhere between $A$ and
$B$. Either of these moves shrinks the effective length of the interval occupied
by the four candidates, defined by
\begin{equation}
L \equiv \max_{i,j} \left( \left\| x_i \right\| - \left\| x_j \right\| \right),
\label{eq:L}
\end{equation}
because $B$ and $a$ have moved inward from the boundary.
On the other hand, $A$'s move to a gap between $a$ and $b$ is length-preserving
in terms of the interval length because $A$ was already inside the interval.
Note that candidate $b$ cannot win no matter
where he or she moves, so $abAB$ has a self-loop, which is omitted in the
diagram for the sake of brevity.

In general, a winner exists in any arbitrary configuration, and he or she wins by keeping the same position. For this reason, each node must have a self-loop, and we will not draw it explicitly. Although each node has four incoming links and four outgoing ones in principle, therefore, only three incoming and three outgoing links are enough to represent the connection structure around each node.
By combining all such elementary transitions, we obtain a graph as shown in
Fig.~\ref{fig:transition}(b), where $L$-decreasing and $L$-preserving
transitions are colored red and black, respectively. At any node of the graph,
it is impossible to increase the effective interval length. If we run a
random walk on this graph, it cannot keep following black arrows, implying that
$L$ will keep shrinking until everyone has no reason to move
further, i.e., arriving at a Nash equilibrium with $L \to 0$.

\section{Summary and Discussion}
In summary, we have investigated the origin of persistent polarization in democracy and proposed an answer based on the existence of multiple political centers. We have constructed a minimalist model of primary elections before the final and showed that all the four candidates keep
the same distance from the median voter at the Nash equilibrium reached by best-response dynamics.
We have found that such a structure tends to intensify political polarization as shown in Fig.~\ref{fig:simul}(b).
This conclusion is consistent with an argument that closed primaries tend to
perpetuate polarization~\cite{wang2021systems}. Possible reforms include
open participation in primaries and a top-two primary system,
although their empirical evidence is not decisive
yet~\cite{mcghee2014primary,mcghee2017has,grose2020reducing}.

As an extension of the MVH, our model inherits its
modeling assumptions of one-dimensionality and single-peakedness
(Sec.~\ref{sec:mvh}), along with its strengths and weaknesses.
Practically speaking, this model implies an informationally
demanding task because all the players
must be able to identify candidates'
positions accurately. Although unrealistic, we stick to this assumption because
our goal has been to propose a mechanism of polarization among such ideal
agents instead of relying on their imperfections.
Another assumption that we have added is that
each candidate's position remains fixed between the two rounds of election.
Of course, candidates could shift back to more centrist positions after the primaries,
but our assumption still describes a part of reality
in the sense that no candidates would want their pledges
to be taken as matters of political expediency.

We have analyzed the resulting two-round election by using the similarity to
the semifinalists' dilemma~\cite{baek2015nash},
and our system can actually be regarded as its limiting case in which only the
final winner is rewarded (Sec.~\ref{sec:semi}).
Based on this observation, some might attribute polarization to the winner-take-all structure in majoritarian elections.
However, a counter-intuitive prediction of our theory is that polarization will only increase further if the payoff for the second-place winner, i.e., the loser of the final election, also becomes valuable enough:
If it definitely pays to secure the victory in
the primary election,
candidates will be motivated to appeal only to their supporters whose median is located far from the political center of the whole electorate.
Moreover, the resulting polarization may exhibit strong hysteresis because
the corresponding strategy profile $(x_1, x_2, x_3, x_4) = (1, 1, 1, 1)$ is a Nash equilibrium
in the semifinalists' dilemma
even for $u=0$. It would be an intriguing future direction to check this
prediction empirically.

\section*{Acknowledgments}
S.K.B. acknowledges support by Basic Science Research Program through the National Research Foundation of Korea (NRF) funded by the Ministry of Education (NRF-2020R1I1A2071670).
H.C.J. was supported by a
National Research Foundation of Korea (NRF) grant funded by the
Korean government(MSIT)(No. NRF-2021R1F1A1063238).


\end{document}